\documentclass[aps,eqsecnum,preprint,prd,showpacs]{revtex4}
\usepackage{graphicx}
\usepackage[latin1]{inputenc}
\usepackage[T1]{fontenc}
\usepackage[english]{babel}
\usepackage{epsfig}
\usepackage{amssymb}
\usepackage[centertags]{amsmath}
\usepackage{amsthm}
\usepackage{amsfonts}
\usepackage{graphics}
\usepackage{graphicx}
\usepackage{psfrag}
\usepackage{subfigure}
\usepackage{dsfont}
\usepackage{eucal}
\usepackage{eufrak}
\usepackage{color}
\usepackage{bm}

\numberwithin{equation}{section}

\def\bold#1{\setbox0=\hbox{$#1$}%
     \kern-.025em\copy0\kern-\wd0
     \kern.05em\%\baselineskip=18ptemptcopy0\kern-\wd0
     \kern-.025em\raise.0433em\box0 }
\def\slash#1{\setbox0=\hbox{$#1$}#1\hskip-\wd0\dimen0=5pt\advance
         to\wd0{\hss\sl/\/\hss}}
\def \de {\partial}

\def \D {\Delta}
\def \b {\beta}
\def \a {\alpha}

\def \g {\gamma}

\def \D {\Delta}
\def \e {\varepsilon}

\def \r {\rho}

\def \m {\mu}
\def \n {\nu}

\def \zh {\hat z}

\def \be {\begin{equation}}
\def \ee {\end{equation}}
\def \bea {\begin{eqnarray}}
\def \eea {\end{eqnarray}}
\def \non {\nonumber}
\def \nn {\nonumber}
\def \noi {\noindent}
\def \ra {\rightarrow}
\def \xra {\xrightarrow}

\def \pr {\prime}

\def \fr {\displaystyle\frac}

\def\laq{~\raise 0.4ex\hbox{$<$}\kern -0.8em\lower 0.62
ex\hbox{$\sim$}~}
\def\gaq{~\raise 0.4ex\hbox{$>$}\kern -0.7em\lower 0.62
ex\hbox{$\sim$}~}

\def \wt {\widetilde}

\begin{document}
\begin{titlepage}
\addtolength{\jot}{10pt}
\addtolength{\jot}{10pt}


\vspace*{1cm}

\title{\bf  Investigating AdS/QCD  duality
through the scalar glueball  correlator \\}

\author{P. Colangelo$^a$, F. De Fazio$^a$, F.~Jugeau$^{b,c}$ and S.~Nicotri$^{a,d}$}

\affiliation{ $^a$ Istituto Nazionale di Fisica Nucleare, Sezione di Bari, Italy\\
$^b$ Institute of High Energy Physics, Chinese Academy of Sciences, Beijing, China\\
$^c$ Theoretical Physics Center for Science Facilities, Chinese Academy of Sciences, Beijing, China\\
$^d$ IPPP, Physics Department, Durham University, UK\\}

\begin{abstract}
We investigate AdS/QCD duality for the two-point correlation
function of the lowest dimension scalar glueball operator, in
the case of the IR soft wall model. We point out the role of the
boundary conditions for the bulk-to-boundary propagator in
determining the gluon condensates. We show how a low energy QCD
theorem can be obtained within the AdS approach, together with a
gluon condensate close to the commonly accepted value  and
robust against perturbation of the background dilaton field.
\end{abstract}

\vspace*{1cm} 

\maketitle
\thispagestyle{empty}
\end{titlepage}

\newpage
\section{Introduction}
The idea that an analytic approach to QCD  in the strong
coupling regime is a viable possibility has recently received
support from the recognition that the AdS/CFT correspondence
conjecture \cite{Maldacena:1997re,Witten:1998qj,Gubser:1998}
could  be applied to QCD-like gauge theories
\cite{Witten2:1998}. In the AdS/CFT conjecture, a correspondence
is supposed to exist between the supergravity limit of a
Superstring/M theory defined in a $AdS_d \times S^{D-d}$
holographic space of total dimension $D$ and the large $N$ limit
of a maximally superconformal  $SU(N)$ YM theory living on the
boundary $\partial AdS_{d-1}$ of dimension $d-1$. Therefore QCD,
being neither supersymmetric nor conformal, is different from
the YM theories to which the AdS/CFT correspondence conjecture
is applied; however, in the regime where its coupling is nearly
constant and where other scales such as the quark masses  can be
neglected, it can be considered as an approximately conformal
field theory defined on the boundary of an AdS-like holographic
space. However, since confinement breaks conformal invariance,
the AdS geometry needs, at least, to be modified to account for
such a phenomenon. Indeed, various models have been proposed,
namely  introducing a maximum value (IR hard wall) of the
$d^{\emph{th}}$ holographic coordinate of the $AdS_d$ space
interpreted as an energy scale (the IR hard wall model
\cite{Polchinski:2001tt,Erlich}), or considering an IR soft wall
described by  a background dilaton field (the IR soft wall model
\cite{andreev0,son2}). In both cases, a mass scale of the order
$\Lambda_{QCD}$ is introduced and  the conformal invariance is
broken. With such modifications of the AdS space, studies have
been carried out in order to check whether properties of QCD are
reproduced in the AdS description; in particular, the meson
spectroscopy and the problem of  chiral symmetry breaking have
been considered, following a procedure which is usually denoted
as the bottom-up approach to  the AdS/QCD correspondence
\cite{AdSQCD,AdSQCD2,AdSQCD3,AdSQCD4,AdSQCD5,AdSQCDa,AdSQCDa1,AdSQCDa2}.

The glueball sector is one of the first issues  investigated
through the AdS/QCD correspondence hypothesis, but mainly in the
(so-called top-down) approach which  attempts to extend the
AdS/CFT duality to QCD-like gauge theories starting from a
theory in a high dimensional space and breaking the conformal
invariance and the supersymmetry, respectively, by
compactification and by appropriate boundary conditions on the
compactified dimensions \cite{Witten2:1998}. In this  way QCD is
recovered as a $4D$ realization of a theory in higher
dimensions, and scalar and tensor glueballs have been
investigated in this framework
\cite{Ooguri,CsakiOoguri,demello,Zyskin,Minahan,ConstableMyers,BrowerMathur,BrowerMathur2,Babington}.
On the other hand, glueballs have also been studied in  the
bottom-up approach
\cite{BoschiFilho:2002vd,BoschiFilho:2002vd2,BoschiFilhoBraga}.
The spectra of scalar and vector glueballs in the IR soft wall
model have been determined, finding that they follow a linear
Regge behaviour; moreover, the effects produced modifying the
AdS geometry or the dilaton profile in the IR have also been
investigated, obtaining that perturbations influence in
different ways the scalar and vector glueball masses
\cite{colangelo}.

Further information can be obtained from the analysis of the
correlation function, as done e.g. in \cite{son2,radyushkin} in
the meson case. In the present work, we are interested in
investigating how the AdS/QCD correspondence is realized for the
two-point correlation functions of  scalar glueball operators.
Similar analyses have been recently presented
\cite{Forkel:2007ru,Beijing}, but our conclusions are different,
mainly as far as the values of the gluon condensates are
concerned, together with their relation to the boundary
conditions imposed in the holographic space.

The plan of the paper is the following. In Section
\ref{sec:twop}, we study the two-point correlation functions of
the lowest dimension QCD operator describing the scalar
glueballs, showing how the condensates are related to the
bulk-to-boundary propagator defined in the holographic space. In
Section \ref{sec:klebanov}, we focus on the dimension four gluon
condensate derived from an AdS/CFT argument worked out by
Klebanov and Witten; this argument consists in relating the
condensate to the expansion near the UV brane of the field
holographically dual to the scalar glueball operator. In Section
\ref{sec:pert}, we investigate how the correlation function is
modified if the background dilaton field is perturbed, and the
way the gluon condensate depends on such a perturbation. At the
end, we present our conclusion.

\section{AdS/QCD duality in the glueball sector}\label{sec:twop}

We consider a five dimensional conformally flat spacetime (the
bulk) described by the metric:
\begin{eqnarray}\label{metric}
g_{MN}&=&e^{2A(z)}\eta_{\,MN} \nn\\
 ds^2&=&e^{2A(z)}(\eta_{\,\mu\nu}dx^\mu
dx^\nu+dz^2)
\end{eqnarray}
with $x^M=(x^\mu,z)$ and $\eta_{\,MN}=\hbox{diag}(-1,1,1,1,1)$.
The coordinates $x^\mu$ ($\mu=0,\dots 3$) are the usual four
dimensional space-time (the boundary) coordinates, while $z$ is
the fifth holographic coordinate running from zero to infinity.
The metric function $A(z)$ satisfies the condition:
\begin{equation}\label{UV}
A(z)\underset{z\rightarrow0}{\rightarrow}\ln
\biggl(\frac{R}{z}\biggr)
\end{equation}
to reproduce the $AdS_5$ metric close  to the UV brane $z \to
0$; $R$ is the anti-de Sitter radius. We adopt the simplest
choice compatible with this constraint:
\begin{equation}\label{UV1}
 A(z)=\ln\biggl(\frac{R}{z}\biggr) \,\,\,.
\end{equation}
In order to softly break the conformal invariance, we consider a
background dilaton field $\phi(z)$ which only depends on the
holographic coordinate $z$ and vanishes at the UV brane. The
large $z$ dependence of this field:
\begin{equation}\label{dilaton}
\phi(z)\underset{z\rightarrow\infty}{\rightarrow} z^2
\end{equation}
is chosen to reproduce the linear Regge behaviour of the
low-lying mesons \cite{son2}. The simplest choice compatible
with the two constraints is
\begin{equation}\label{dilaton1}
\phi(z) = c^2 z^2
\end{equation}
and involves the dimensionful parameter $c$ which sets the scale
of QCD quantities, such as the hadron masses.

In order to investigate the glueball sector of QCD, we consider
the lowest dimension QCD operator having the scalar glueball
quantum numbers:
$\CMcal{O}_{S}=\beta(\a_s)\,F_{\m\n}^a\,F^{\m\n\,a}$ ($a=1,\dots
8$ a color index) with $J^{PC}=0^{++}$ and conformal dimension
$\Delta=4$; $\beta(\a_s)$ is the Callan-Symanzik function. This
operator is defined in the field theory living on the $4d$
boundary. According to the AdS/CFT correspondence, the conformal
dimension of a ($p$-form) operator on the boundary is related to
the $(AdS\;mass)^{2}$ of the dual field in the bulk by the
relation \cite{Witten:1998qj}:
\begin{equation}
(AdS\;mass)^2R^2=(\Delta-p)(\Delta+p-4)\;.\label{m5}
\end{equation}
We assume that the mass $m_{5}^{2}$ of the scalar bulk field
$X(x,z)$ constructed as the holographic correspondent of the
operator $\CMcal{O}_S$ is given by this expression. It is described
by the action with the gravity and the dilaton background:
\begin{equation}
S_{5d}^{eff}=-\frac{1}{2\,k}\int d^5x\,\sqrt{-g}\,e^{-\phi(z)}\,\,
g^{\,MN}(\partial_{M}X) (\partial_{N}X)
\,\,\,\,\label{actionscalmass}
\end{equation}
with $g=det(g_{MN})$ and $k$ a dimensionful parameter. As
discussed in \cite{colangelo}, scalar glueball states can be
identified as the normalizable modes of $X(x,z)$ satisfying the
equations of motion obtained from \eqref{actionscalmass},  and
their mass spectrum can be obtained consequently.

In the following, we consider the two-point correlation function
of the QCD operator $\CMcal{O}_{S}$:
\begin{equation}\label{QCDcorrelator}
\Pi_{QCD}(q^2)=i\int d^4 x \, e^{iq\cdot x}\langle0|T\big[\CMcal{O}_{S}({x})\CMcal{O}_{S}(0)\big]|0\rangle\; .
\end{equation}
In QCD, the short-distance expansion of this correlation
function (obtained in our metric for $q^2>0$), writing
$\beta(\alpha_s)=\beta_1 \left( \frac{\alpha_s}{\pi}\right)+
\beta_2 \left( \frac{\alpha_s}{\pi}\right)^2 +\dots $ and
keeping the first term with
$\b_{1}=-\frac{11}{6}N_{c}+\frac{1}{3}n_{F}$ ($N_{c}$ and
$n_{F}$ are, respectively, the number of colors and of active
flavours; we use $n_F=0$), can be expressed in terms of a
perturbative contribution and a series of power corrections; the
result, which includes the leading order perturbative term and
the power corrections up to  $1/q^4$, reads as
\cite{Novikov,Novikov2,Paver,Paver2,Narison}:
\begin{eqnarray}\label{QCD2pt}
  \Pi_{QCD}(q^2)&=&C_{0}\,q^4\biggl(-\ln{\big(\fr{q^2}{\n^2}\big)}+2-\fr{1}{\e^{\pr}}\biggr)
  \nonumber \\&+&C_{4}\langle\CMcal{O}_4\rangle
  +\frac{C_6}{q^2}\langle\CMcal{O}_6\rangle+\frac{C_8}{q^4}\langle\CMcal{O}_8\rangle\; ,
\end{eqnarray}
$\n$ being a renormalization scale. Eq.~\eqref{QCD2pt} involves
the coefficients:
\begin{eqnarray}
C_{0}&=&2\biggl(\fr{\b_1}{\pi}\biggr)^2\biggl(\fr{\a_s}{\pi}\biggr)^2 \non \\
C_{4}&=&4\b_1^2\,\Big(\frac{\a_{s}}{\pi}\Big) \\
C_6&=&8\b_1^2\Big(\frac{\a_s}{\pi}\Big)^2  \non \\
C_8&=&8\pi\Big(\frac{\b_1}{\pi}\Big)^2\a_s^3  \non
\end{eqnarray}
and the QCD condensates of dimension four, six and eight, respectively:
\begin{eqnarray}
\langle\CMcal{O}_4\rangle&=&\langle \frac{\alpha_s}{\pi} \,F_{\m\n}^a\,F^{\m\n a}\rangle   \non \\
\langle\CMcal{O}_6\rangle&=&\langle g_s f_{abc}\,F_{\m\n}^a\,F_{\n\rho}^b\,F_{\rho\m}^c\rangle \\
\langle\CMcal{O}_8\rangle&=&14\langle\Big(f_{abc}\,F_{\m\a}^a\,F_{\n\a}^b\Big)^2\rangle-\langle\Big(f_{abc}\,F_{\m\n}^a\,F_{\a\b}^b\Big)^2\rangle\;.  \non
\end{eqnarray}

In addition to the short-distance expansion (in which, however,
possible instanton contributions have not been considered), a
Low Energy Theorem is known  for the correlation function
\eqref{QCDcorrelator} \cite{Novikov,Novikov2}:
\begin{equation}\label{pi2}
 \Pi_{QCD}(0)=-16 \beta_1 \langle \CMcal{O}_4\rangle \,\,\,,
\end{equation}
a theorem relating the value of the correlation function at zero momentum to the dimension four
gluon condensate.

Within the AdS/CFT correspondence, the correlation function
\eqref{QCDcorrelator} can be computed using the equivalence
between the generating functional of the connected correlators
in the $4d$ theory and the effective action in the $5d$ bulk
theory. The conjecture is \cite{Witten:1998qj,Gubser:1998}:
\begin{equation}\label{generating}
  \biggl\langle e^{i\int d^4 x\;X_0(x )\,\CMcal{O}_{S}(x)}\biggr\rangle_{CFT}=
  e^{iS_{5d}^{eff}[X(x,z)]}
\end{equation}
where $S_{5d}^{eff}[X(x,z)]$ is the effective (classical) action
\eqref{actionscalmass} of the scalar bulk field $X(x,z)$ dual to the
operator $\CMcal{O}_{S}(x)$, for which $X_0(x)$ is the source
defined  on the boundary. The relation between the field $X(x,z)$
and its boundary value $X_0(x)$ is implemented by defining the
bulk-to-boundary propagator $K$:
\begin{equation}\label{BC}
X(x,z)=\int
d^4 x^{\pr}\,K(x-x^\prime;z,0)\,X_0(x^\prime)\; .
\end{equation}
To achieve the condition
$X(x,z)\xrightarrow[z\rightarrow0]{}X_0(x)$, the
bulk-to-boundary propagator $K$ must reduce to a delta function
on the boundary:
\begin{equation}\label{deltalimit}
K(x-x^\prime;z,0)\xrightarrow[z\rightarrow0]{}\delta^4(x-x^\prime)\;.
\end{equation}
In the Fourier space, the convolution \eqref{BC} is replaced by
\begin{equation}\label{BC fourier}
  \wt{X}(q,z)=\wt{K}(q^2,z)\,\wt{X}_0(q)\;
\end{equation}
and involves the $4d$  Fourier transforms (with respect to $x^\m$)
$\tilde X$, $\tilde X_0$ and $\tilde K$ of $X$, $X_0$ and $K$, respectively.

Furthermore, an AdS/CFT prescription also requires the vanishing
of the bulk-to-boundary propagator in the deep IR region:
\begin{equation}
K(x-x';z,0)\xrightarrow[z\rightarrow\infty]{}0\;\,\,\, ;
\end{equation}
we discuss  this IR boundary condition in the following.

The equation of motion for the field $X(x,z)$ obtained from the
action \eqref{actionscalmass} is:
\begin{equation}\label{EOM}
  \eta^{MN}\de_M\biggl[e^{-B(z)}\,\de_N X(x,z)\biggr]=0
\end{equation}
where $B(z)$ is a combination of the metric function $A(z)$ and
of the dilaton background field $\phi(z)$: $B(z)=\phi(z)-3
A(z)$. It can be written as an equation for the  Fourier
transform of the bulk-to-boundary propagator $\wt{K}(q^2,z)$:
\begin{eqnarray}\label{equationK}
  e^{B(z)}\de_z\biggl[e^{-B(z)}\de_z
  \wt{K}(q^2,z)\biggr]-q^2\wt{K}(q^2,z)&=&0\;.
\end{eqnarray}
From now on, we use the dimensionless variable $\zh=c z$. The
general solution of this equation can be written as:
\begin{equation}\label{general integral}
  \wt{K}\Big(\frac{q^2}{c^2},\zh^2\Big)=\wt{A} \, \wt{K}_1\Big(\frac{q^2}{c^2},\zh^2\Big)+\wt{B} \, \wt{K}_2\Big(\frac{q^2}{c^2},\zh^2\Big) \,\, ;
\end{equation}
$\wt{A}$ and $\wt{B}$, constants with respect to $\zh$, depend
in general on  $q^2/c^2$. In order to determine the two
independent solutions $\wt{K}_{1,2}$,  it is convenient to solve
the equation for the Bogoliubov transform of $\wt{K}$:
\begin{equation}
 \wt{Y}= e^{-B(\zh)/2} \wt{K} \,\,\, ;
\end{equation}
one obtains a one-dimensional Schr\"odinger-like equation
\cite{colangelo}:
\begin{equation}\label{eqbogol}
 - \wt{Y}''\Big(\frac{q^2}{c^2},\zh^2\Big)+V(\zh)\wt{Y}\Big(\frac{q^2}{c^2},\zh^2\Big)=0
\end{equation}
where the derivatives act on $\zh$ and  the potential is:
\begin{equation}\label{potential}
  V(\zh)= \zh^2+\frac{15}{4\zh^2}+2+\frac{q^2}{c^2} \,\,\, .
\end{equation}
Two independent solutions $ \wt{K}_1$ and $ \wt{K}_2$ of
\eqref{eqbogol} are:
\begin{equation}
  \wt{K}_1\Big(\frac{q^2}{c^2},\zh^2\Big)  =  U\left(\fr{q^2}{4c^2},-1, \zh^2\right)
  \label{LaguerreTricomi1}
\end{equation}
with $U$ the Tricomi confluent hypergeometric function, and
\begin{equation}
  \wt{K}_2\Big(\frac{q^2}{c^2},\zh^2\Big)  = L\left(-\fr{q^2}{4c^2},-2, \zh^2\right)
  \label{LaguerreTricomi2}
\end{equation}
where $L$ is the generalized Laguerre function. The function
$\wt{K}_2$ vanishes  as $\zh^4$ when $\zh \ra0$, while
$\wt{K}_1$ has a finite value in this limit:  the UV boundary
condition \eqref{deltalimit} imposes that
$\wt{A}=\Gamma(2+\frac{q^2}{4 c^2})$ in \eqref{general
integral}, so that the behaviour
\begin{equation}
  \wt{K}\Big(\frac{q^2}{c^2},\zh^2\Big)\xrightarrow[\zh\rightarrow0]{}1\;
\end{equation}
is recovered for any value of $q^2$. The low-$z$
expansion of $\wt{K}$ reads:
\begin{eqnarray}\label{kzto0}
  \wt{K}(q^2)=1-\fr{q^2z^2}{4}&+&\fr{1}{64}\biggl[32\wt{B}\,c^4 - (4\g_E-3)\,q^2(q^2+4c^2) \nn \\
  &&-2q^2(q^2+4c^2)\ln(c^2z^2)-2q^2(q^2+4c^2)\psi(2+\fr{q^2}{4c^2})\biggr]z^4+{O}(z^6) \,\,\, . \nn \\
\end{eqnarray}
On the other hand, the behaviour at infinity of $\tilde{K}_1$
and $\tilde{K}_2$ in \eqref{LaguerreTricomi1} and
\eqref{LaguerreTricomi2} is very different. For $q^2<0$
$\tilde{K}_1$ diverges when $\zh\ra\infty$, while it goes to
zero for space-like four-momenta. $\tilde{K}_2$ has a strong
divergence in the IR  for any value of  $q^2$. A possible IR
boundary condition in the soft wall model is that the effective
action is finite in this region \cite{son2}. According to this
prescription, $\tilde{K}_2$ should be discarded in the
calculation of \eqref{generating} (that is, $\wt{B}=0$ in
\eqref{general integral}). However, as we discuss below,  this
requirement seems to produce inconsistencies in the two-point
correlation function; therefore, it is worth investigating the
possible role of contributions (when $z\ra0$) brought by
$\wt{K}_2$ to the AdS version of the correlation function.  This
means that the effective action in the bulk requires a
regularization in the IR.

We evaluate the correlator following the AdS/CFT correspondence
relation \eqref{generating}, which implies
\cite{test,test2,test3,test4,test5,test6,test7,test8}:
\begin{eqnarray}\label{generating derivative}
  i\int d^4 x \,e^{iq\cdot
  x}\,\frac{(-i)^2\delta^2}{\delta X_0(x)\delta X_0(0)}
  \biggl\langle e^{i\int d^4 x^\pr \;X_0(x^\pr)\,
  \CMcal{O}_S(x^\pr)}\biggr\rangle_{CFT}\Big|_{X_{0}=0}= i\int d^4
  x\,e^{iq\cdot x}\,\frac{(-i)^2i\delta^2S_{5d}^{eff}}{\delta
  X_{0}(x)\delta X_{0}(0)}\Big|_{X_{0}=0}\;. \nn \\
\end{eqnarray}

Deriving twice eq.\eqref{generating}, we obtain in terms of the
variable $\zh$:
\begin{eqnarray}
 \Pi_{AdS}(q^2)  = -
\Big(\frac{R^3}{2\,k}\Big)c^4\int
d^4q^{\pr}\,\delta^4(q+q^{\pr})\frac{e^{-\zh^2}}{\zh^3}
\biggl[ &\wt{K}&\Big(\frac{q'\,^2}{c^2},\zh^2\Big)\partial_{\zh}\wt{K}\Big(\frac{q^2}{c^2},\zh^2\Big)\nn \\
&+&\wt{K}\Big(\frac{q^2}{c^2},\zh^2\Big)
\partial_{\zh}\wt{K}\Big(\frac{q'\,^2}{c^2},\zh^2\Big)\biggr]\biggl|_{\zh \to0}^{\zh \to \infty} \nn \\
\end{eqnarray}
so that the integration over $q^{\pr}$ gives the AdS
representation of two-point correlation function in terms of the
bulk-to-boundary propagator:
\begin{equation}\label{AdS correlator}
  \Pi_{AdS}(q^2)=-\frac{R^3}{k}c^4\wt{K}\Big(\frac{q^2}{c^2},\zh^2\Big)\frac{e^{-\zh^2}}{\zh^3}\partial_{\zh}\wt{K}\Big(\frac{q^2}{c^2},\zh^2\Big)\biggl|_{\zh\ra0}^{\zh\ra \infty}\;.
\end{equation}
Using the solution \eqref{general integral} the expression
\eqref{AdS correlator} is singular both in the UV and in the IR.
A regularization prescription consists in considering a new
effective action:
\begin{equation}\label{redefinition}
  S_{reg.}^{eff}=S^{eff}_{5d}-S^{eff}_{c.t.}\Big|_{z=\epsilon}-S^{eff}_{c.t.}\Big|_{z=\Lambda}
\end{equation}
where the two  terms are introduced to subtract the UV (when
$z=\epsilon\to0$) and IR (when $z=\Lambda \to \infty$)
divergences; the first one is the usual term considered in the
AdS/CFT procedure, while the second one defines the IR soft wall
model: it involves $\tilde B$, and vanishes when $\tilde B=0$ as
in the usual procedure. The expression of the two-point
correlation function reads now:
\begin{equation}\label{AdS correlator1}
  \Pi_{AdS}(q^2)  =  \fr{R^3}{8 k}\biggl\{2 \wt{B} \,c^4
  -q^2(q^2+4c^2)\Big(\ln(c^2 \epsilon^2)+\psi{\left(2+\fr{q^2}{4c^2}\right)}   +\g_E-3 \Big)\biggr\} \,\,\,
\end{equation}
where $\psi{(x)}$ is the Euler function, the logarithmic
derivative of $\Gamma$. This AdS representation of the two-point
correlation function\eqref{QCDcorrelator} should match the QCD
result.

Let us consider the short-distance regime. Expanding \eqref{AdS
correlator1} for $q^2\to+\infty$:
\begin{eqnarray}\label{AdS correlator2}
  \Pi_{AdS}^{reg}(q^2) & = &
  \fr{R^3}{k}\biggl\{q^4\cdot\fr{1}{8}\left[2-2\g_E+\ln4-\ln(q^2 \epsilon^2)\right] \nn \\
  &&+q^2\left[\frac{c^2}{2}\ln(q^2 \epsilon^2)+\fr{c^2}{4}\left(1-4\g_E+2\ln4\right)\right]+ \\
  && +\fr{c^4}{6}\,(12 \wt{B}-5)
  + \fr{2c^6}{3}\fr{1}{q^2}-\fr{4c^8}{15}\fr{1}{q^4}+{O}\left(\fr{1}{q^6}\right)\biggr\}\;\;, \nn
\end{eqnarray}
and identifying $\epsilon$ with the renormalization scale
$\n^{-1}$ (as in \cite{csaki}, since holography is related to
the renormalization group
\cite{verlinde,BianchiFreedmanSkenderis}), we obtain:
\begin{eqnarray}\label{AdS correlator3}
  \Pi_{AdS}(q^2) & = & \fr{R^3}{k}\biggl\{q^4\cdot\fr{1}{8}\left[2-2\g_E+\ln4-\ln(\frac{q^2}{\n^2})
  \right] \nn \\
  &&+q^2\left[-\frac{c^2}{2}\ln(\frac{q^2}{\n^2})+\fr{c^2}{4}\left(1-4\g_E+2\ln4\right)\right]+ \\
  && +\fr{c^4}{6}\,(12 \wt{B}-5) + \fr{2c^6}{3}\fr{1}{q^2}-\fr{4c^8}{15}\fr{1}{q^4}+{
  O}\left(\fr{1}{q^6}\right)\biggr\}\;\;. \nn
\end{eqnarray}

We can now compare the QCD \eqref{QCD2pt} and the AdS \eqref{AdS
correlator3} expressions: if any mismatch occurs, duality is
violated. Identifying the $q^4 \ln(q^2)$ terms in \eqref{QCD2pt}
and \eqref{AdS correlator3} we get a condition which fixes the
$k$ parameter in the action \eqref{actionscalmass}:
\begin{equation}\label{kmatch}
  k=\frac{\pi^4}{16\a_s^2 \beta_1^2}R^3 \,\,\,.
\end{equation}
This condition does not involve the parameter $c$, related to
the dilaton background, but is only connected to the AdS
structure of the holographic space-time close to the UV
boundary.

The other terms in the power expansion of the two-point
correlation function involve the various gluon condensates and
deserve discussion. Without the contribution of $\wt{K}_2$ (i.e.
for $\tilde B=0$), the dimension four gluon condensate obtained
comparing \eqref{QCD2pt} and \eqref{AdS correlator3}, is given
by:
\begin{equation}\label{negativegcond}
\langle\frac{\a_{s}}{\pi}\,F^2\rangle=-\frac{10\,\a_s}{3\pi^3}c^4\;;
\end{equation}
it turns out to be negative, as also observed in
\cite{Forkel:2007ru}. A negative value is not compatible with
current determinations of this QCD quantity \cite{Reinders,Reinders2,khodjamirian}.
Moreover, it would be inconsistent with a positive value
obtained analyzing the two-point correlation function  of vector
operators \cite{Beijing}, and with a determination based on
computing the Wilson loop in the AdS/QCD approach  in which the
value $\langle\frac{\a_{s}}{\pi}\,F^2\rangle=0.010\pm0.0023$
GeV$^4$ was found \cite{andreev}. In both the calculations,   a
dilaton background field was chosen to break  the conformal
invariance, i.e. the IR soft wall model was used.

Another observation concerns the two-point correlation function
at zero momentum. From \eqref{AdS correlator1} we obtain if
$\tilde B=0$:
\begin{equation}\label{LET}
\Pi_{AdS}(0)=0\;\;,
\end{equation}
while the Low Energy Theorem  \eqref{pi2} relates $\Pi_{QCD}(0)$
to the dimension four gluon condensate. These shortcomings could
cast doubts on the approach based on the IR soft wall model.
However, they can be removed if the contribution of the solution
$\wt{K}_2$ in \eqref{general integral} is  taken into account in
the determination of the AdS version of the two-point
correlation function: if we allow $\wt{K}_2$ to contribute to
the correlation function, it seems possible to recover AdS/QCD
duality.

Let us discuss this point. The coefficient function $\wt{B}(\frac{q^2}{c^2})$ is
undetermined at this stage. In general, $\wt{B}(\frac{q^2}{c^2})$
can be any function of the dimensionless ratio $q^2/c^2$. If
$\wt{B}(\frac{q^2}{c^2})$ is a function having the polynomial
behaviour at large space-like $q^2$, namely:
\begin{eqnarray}\label{expression}
  \wt{B}\left(\frac{q^2}{c^2}\right) &
  \xra[q^2\ra+\infty]{}&\eta_{1}\frac{q^2}{c^2} +\eta_{0}\;,
\end{eqnarray}
its parameters can be fixed by matching the AdS expression of
the two-point correlation function with the OPE expansion
\eqref{QCD2pt} of $\Pi_{QCD}(q^2)$. A constant term $\eta_0$ in
$\wt{B}$ contributes to term involving the the dimension four
gluon condensate in the power expansion, so that the $D=4$ gluon
condensate reads:
\begin{equation}\label{gcond}
\langle\frac{\a_{s}}{\pi}\,F^2\rangle=\frac{4\a_s}{\pi^3}\Big(2\eta_0-\frac{5}{6}\Big)c^4 \,\,\, ,
\end{equation}
while the other two condensates are given by
\begin{eqnarray}\label{highcond}
\langle\CMcal{O}_6\rangle&=&\frac{4}{3\pi^2}c^6 \nn\\
\langle\CMcal{O}_8\rangle&=&-\frac{8 }{15\alpha_s \pi^3}c^8 \,\,\,
\,\, . \label{highcond1}
\end{eqnarray}
Notice that the $c$ parameter in the dilaton field sets the
scale for all the condensates.

The coefficient $\eta_0$ in \eqref{gcond}  can be fixed using the
commonly used  numerical value  of the condensate
\cite{Reinders,Reinders2,khodjamirian}:
\begin{equation}\label{gcond1}
\langle\frac{\a_{s}}{\pi}\,F^2\rangle\simeq 0.012 \,\, {\rm GeV}^4
\end{equation}
together with the value of the dilaton parameter $c$ from the
spectrum of the $\r$ mesons \cite{son2}: $m_{\r_n}^2=4(n+1)
c^2$. For $\alpha_s=1$ one obtains $\eta_0\simeq2.5$. On the
other hand, the AdS expressions of the dimension six
\eqref{highcond} and dimension eight \eqref{highcond1} gluon
condensates are different in size (and in sign for
$\langle\CMcal{O}_8\rangle$) from their commonly used values
\begin{eqnarray}\label{numval}
\langle\CMcal{O}_6\rangle&\simeq & \,\,\, 0.045 \,\, {\rm GeV}^6 \nn \\
\langle\CMcal{O}_8\rangle&\simeq& \frac{9}{16}
\left(\frac{\pi}{\alpha_s}\right)^2
\left(\langle\frac{\a_{s}}{\pi}\,F^2\rangle \right)^2 \,\,\,
\,\,
\end{eqnarray}
(adopting  in the last expression the factorization
approximation). However, while the estimated uncertainty for the
dimension four condensate is about $30\%$, the values of the
dimension six and eight gluon condensates are very uncertain, so
that the difference between \eqref{highcond} and \eqref{numval}
could be attributed to the poor knowledge of the gluon vacuum
matrix elements.

In the AdS expression of the two point correlator \eqref{AdS
correlator3}, a term $q^2 c^2$ involving a dimension two
condensate appears, as also discussed in
\cite{Forkel:2007ru,Beijing} (see \cite{andreev0} in the case of
the vector current). This term is missing in the QCD
short-distance expansion (in general, the existence and the
meaning of a dimension two condensate, which cannot be expressed
as the vacuum expectation value of a local gauge invariant QCD
operator, is a matter of discussion; a recent analysis of  its
interpretation can be found in \cite{zakharovmartina}). However,
this term could be removed if
\begin{equation}\label{eta2}
\eta_{1}=-\frac{1}{2}\left(\frac{1}{4} -\gamma_E +\ln{2} \right)\;
\end{equation}
and the $\log(z)$ term is included among the contact terms, an
admittedly  fine-tuning procedure.

\begin{figure}[t]
\begin{center}
\vspace*{0.2cm}
\includegraphics[width=0.40\textwidth] {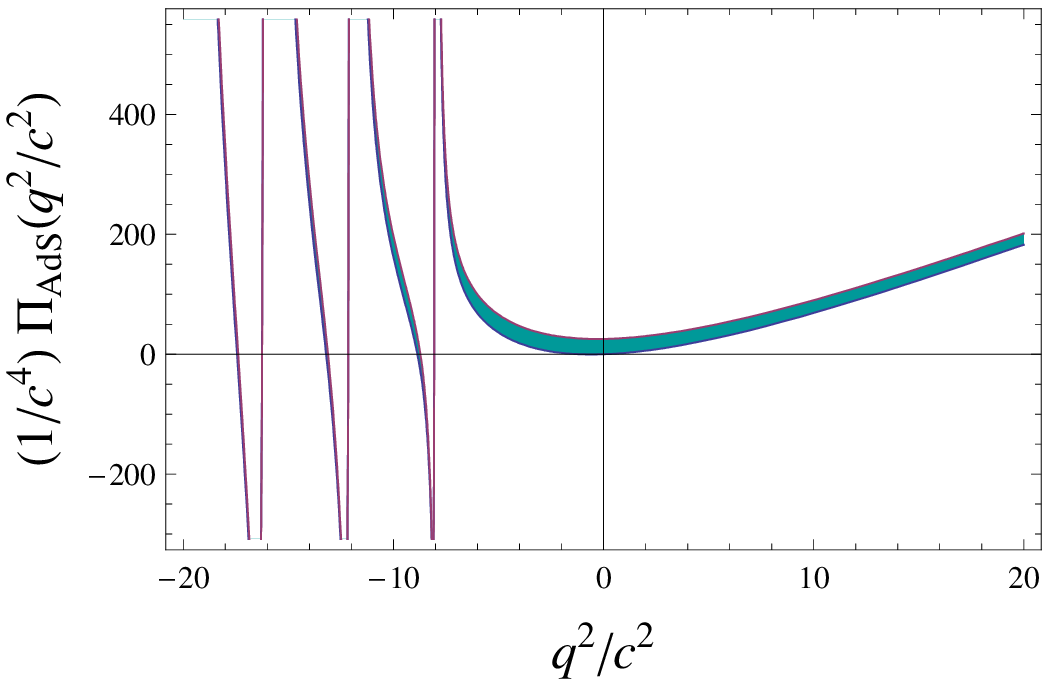} \hspace*{0.5cm}
\includegraphics[width=0.38\textwidth] {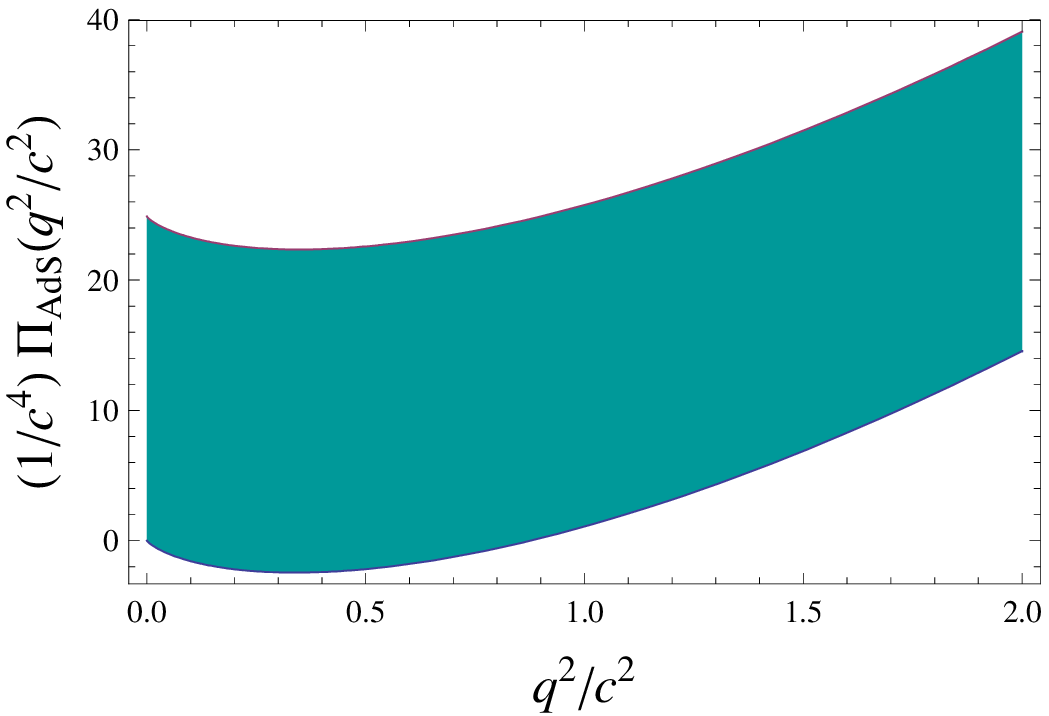}
\end{center}
\caption{\baselineskip=14pt  The two-point correlation function
$\frac{1}{c^4} \Pi_{AdS}(q^2/c^2)$ in \eqref{AdS correlator1}. The
renormalization scale is fixed to $\nu=1$ GeV. The left
panel shows the presence of poles at $q^2_n=-(4n+8)c^2$,
corresponding to the spectrum of the scalar glueballs
\cite{colangelo}. The right panel shows, enlarged,   the range close
to $q^2=0$:  the effect of modifying the gluon condensate, from
$\eta_0=0$ (bottom curve) to  $\eta_0=2.5$ (top curve) is evident. }
\label{fig:plot1}
\end{figure}

Let us now consider the AdS two-point correlation function
\eqref{AdS correlator1} in the full $q^2$ range. In Fig.
\ref{fig:plot1} we plot $\Pi_{AdS}(q^2)$ for two values of
$\eta_0$: $\eta_0=0$ and $\eta_0=2.5$, fixing $\eta_1=0$ and the
scale $\nu=1$ GeV. In the time-like ($q^2<0$) region,  a
discrete set of poles appears according to the spectral relation
($n$ is an integer)
\begin{equation}\label{spectrum}
 m_n^2=(4 n +8)c^2
\end{equation}
obtained in  \cite{colangelo}. The residues
\begin{equation}\label{residues}
  f_n^2= \frac{R^3}{k} 8 (n+1)(n+2)c^6,
\end{equation}
related to the matrix elements $\langle 0| {\CMcal O}_S|
n\rangle$,  coincide with the result obtained in
\cite{Forkel:2007ru}. The structure in poles  in \eqref{AdS
correlator1} is due to  the Euler function
$\psi(2+\frac{q^2}{4c^2})$.  It does not depend on $\wt{B}$, but
it comes from the solution $\wt{K}_1$ in \eqref{general
integral} which can be written according to the spectral
representation \cite{Beijing}:
\begin{equation}\label{analytical}
\wt{A}\wt{K}_1(\frac{q^2}{c^2},\zh^2)=\sqrt{\frac{k}{R^3}}\frac{1}{c}\sum_{n=0}^{\infty}\frac{f_n\;\wt{K}_n(\zh^2)}{q^2+m_n^2+i\epsilon}
\end{equation}
where $\wt{K}_n(\zh^2)$ are  the normalizable modes computed in \cite{colangelo}:
\begin{equation}\label{eigenfunctions}
  \wt{K}_n(\hat{z}^2)=A_n\,\hat{z}^4\,_1F_1\left(\frac{q^2}{4c^2}+2\,,3\,,\hat{z}^2\right)\;.
\end{equation}
Nevertheless, the expression \eqref{analytical} in terms of pole
masses $m_n^2$ and residues $f_n^2$ does not define
unambiguously the general solution \eqref{general integral}
which can indeed contain extra regular terms in $q^2/c^2$, for
example if  $\wt{B}\wt{K}_2$ is a polynomial as in
\eqref{expression}; however, these terms can affect the power
expansion and the value of  the condensates, as discussed, e.g.,
in \cite{Novikov,Novikov2}.

In the space-like region ($q^2>0$)  the behaviour of
$\Pi_{AdS}(q^2)$ is smooth. In the region close to $q^2=0$, as
shown in Fig. \ref{fig:plot1} (right panel),  one finds from
\eqref{AdS correlator1}:
\begin{equation}\label{pi}
 \Pi_{AdS}(0)=\frac{R^3}{k} 2 \wt{B}(0) c^4\,\,\,.
\end{equation}
Calling  $\wt{B}(0)=\wt{\eta}_0$ and using \eqref{gcond}  we
have:
\begin{equation}\label{pi1}
 \Pi_{AdS}(0)=\frac{\alpha_s}{4 \pi} (-\beta_1) \frac{\wt{\eta}_0}{\eta_0-\frac{5}{12}}
\left( -16 \beta_1 \langle\frac{\a_{s}}{\pi}\,F^2\rangle \right)\,\,\,,
\end{equation}
an expression similar to the Low Energy Theorem  \eqref{pi2}.
For a constant coefficient function $\wt{\eta}_0=\eta_0$,  it is
possible to constrain the value of $\eta_0$ imposing that
\eqref{pi1} and \eqref{pi2} coincide:
\begin{equation}
 \eta_0=\frac{5}{12} \left(\frac{1}{1+\frac{\alpha_s}{4 \pi} \beta_1} \right)\,\,\, .
\end{equation}
Using this expression in \eqref{gcond}, together with
$\alpha_s=1$ we find:
\begin{equation}\label{gcondvalue1}
\langle\frac{a_{s}}{\pi}\,F^2\rangle\simeq0.002\;{\rm GeV}^4\;,
\end{equation}
while for $\alpha_s=1.5$ we have:
\begin{equation}\label{gcondvalue2}
\langle\frac{\a_{s}}{\pi}\,F^2\rangle\simeq0.007\;{\rm GeV}^4.
\end{equation}
These values are smaller but close to \eqref{gcond1} and
compatible with the result obtained studying the Wilson loop
\cite{andreev} which indicates that the solution
\eqref{LaguerreTricomi2} could play a role in reconstructing a
bulk-to-boundary propagator which implements the AdS/QCD duality
in this case.

We close this Section observing that  it is possible to
investigate the role of the two solutions analogous to
\eqref{general integral} in the two-point correlation function
of the $J^{PC}=1^{--}$ meson operators  \cite{son2,Beijing} ,
studying duality  between the QCD and the AdS expressions of the
correlation function. The perturbative term  is reproduced: it
does not depend on the parameter $c$. The  dimension four gluon
condensate is positive \cite{Beijing}, and  the effect of
including the subleading solution with a constant coefficient
function $\tilde B$  only consists in modifying the dimension
two condensate.  In this case, however, the contribution of the
dimension six condensate  in the power expansion of $\Pi_{QCD}$
is not reproduced in $\Pi_{AdS}$ if $\tilde B$ is a polynomial
in $q^2/c^2$.

\section{Gluon condensate from an AdS/CFT  argument applied to the glueball sector}\label{sec:klebanov}

It is interesting to consider the dimension four gluon
condensate from another viewpoint. From AdS/CFT correspondence,
a scalar bulk field $X(x,z)$ is required to satisfy  the  UV
boundary condition \cite{Klebanov}:
\begin{equation}\label{zerolimit}
  X(x,z)\xra[z\ra0]{}z^{d-\D}\bigl[X_0(x)+O(z^2)\bigr]+z^\D\bigl[A(x)+{ O}(z^2)\bigr]\;;
\end{equation}
$d$ is the dimension of the boundary space-time and $\Delta$ is
the conformal dimension of the operator associated to the source
field $X_{0}$. $A(x)$ can be considered  as a physical
fluctuation  and is argued to be related to the vacuum
expectation value of the corresponding CFT operator, as obtained
by Klebanov and Witten:
\begin{equation}
  A(x)=\fr{1}{2\D-d}\;\langle\CMcal{O}(x)\rangle \,\,\, .
\end{equation}
In our case,  $d=\Delta=4$ and eq.\eqref{zerolimit} becomes:
\begin{equation}
  X(x,z)\xra[z\ra0]{}[X_0(x)+{ O}(z^2)]+z^4\bigl[A(x)+{ O}(z^2)\bigr]
\end{equation}
\noi with
\begin{equation}
  A(x)=\fr{1}{4}\;\langle \CMcal{O}_S(x)\rangle\;.
\end{equation}
Following this AdS/CFT guideline, let us see if it is possible
to recover the four dimension gluon condensate. In our case, in
the Fourier space, this term appears in the bulk-to-boundary
propagator \eqref{kzto0}, considering the whole contribution
proportional to $z^4$, taking the limit $q^2 \to \infty$ and
keeping the piece independent of $q^2/c^2$. After having
properly renormalized the bulk field $X(x,z)$ in
\eqref{actionscalmass} by absorbing the $1/k$ factor, we get:
\begin{equation}
 \langle \frac{\alpha_s}{\pi} F^2 \rangle =
  \frac{4\a_s}{\pi^2}  \left(2 \eta_0-\frac{5}{6}\right)c^4 \,\,\, .
\end{equation}
The expression is identical to \eqref{gcond}  but for a $1/\pi$
overall factor which could be related to the normalization of
the $A$ term, a shortcoming deserving further investigations in
order to gain a deeper understanding of the AdS/QCD duality.
Notice that, with respect to AdS/CFT, the new aspect is the
presence of the dimensionful parameter $c$, so that terms as
$c^2 z^2$, $c^4 z^4$, etc.,  appear in the expansion
\eqref{zerolimit}.

\section{AdS/QCD duality in a deformed AdS-dilaton background}\label{sec:pert}

The form of the dilaton background field \eqref{dilaton1} is the
simplest choice allowing to fulfill the constraints
\eqref{dilaton} and $\phi(0)=0$. Other expressions could be
used, and indeed in ref.\cite{colangelo} we investigated the
effects on the glueball spectrum of a perturbation of the
dilaton field which, instead of being given by
eq.\eqref{dilaton1}, involves an additional term which does not
spoil the UV ($z\to 0$) and IR ($z \to +\infty$) behaviour:
\begin{equation}\label{dilatonpert}
\phi(z)=c^2 z^2+ \lambda c z \,\,
\end{equation}
with $\lambda$ a small dimensionless parameter. This new
background field does not sensibly modify the Regge behaviour of
the spectrum.

The equation of motion \eqref{EOM} for the bulk-to-boundary
propagator, with the deformed AdS-dilaton background, takes the
form:
\begin{equation}
\wt{K}^{\pr\pr}\left(\frac{q^2}{c^2},\lambda,\zh\right)
+P(\lambda,\zh)\wt{K}^{\pr}\left(\frac{q^2}{c^2},\lambda,\zh\right)+Q(\lambda,\zh)\wt{K}
\left(\frac{q^2}{c^2},\lambda,\zh\right)=0
\end{equation}
where
\begin{eqnarray}
P(\lambda,\zh)&=&-\Big(2\zh+\frac{3}{\zh}+\lambda\Big)\\
Q(\lambda,\zh)&=&-\frac{q^2}{c^2} \,\,\, .
\end{eqnarray}
The Frobenius method allows to obtain the new bulk-to-boundary
propagator:
\begin{equation}\label{general integralpert}
\wt{K}\left(\frac{q^2}{c^2},\lambda,\zh\right)={A}^{\prime}
\wt{K}_1\left(\frac{q^2}{c^2},\lambda,\zh\right)+{B}^{\prime}
\wt{K}_2\left(\frac{q^2}{c^2},\lambda,\zh\right)
\end{equation}
as an expansion of the solutions in $\zh$; ${A}^{\prime}=-4a_0$,
with $a_0$ a real parameter, and $\wt{K}_{1,2}$ given by:
\begin{eqnarray}\label{Kexp}
\wt{K}_1\left(\frac{q^2}{c^2},\lambda,\zh\right)&=&a_0^{-1}\Big(-\frac{1}{4}+A_2\,\zh+B_2
\,\zh^2+C_2\,\zh^3+D_2\,\zh^4
+E_2\,\zh^4\,\ln(\zh)+{O}(\zh^5,\zh^5\ln(\zh))\Big) \nn\\
\wt{K}_2\left(\frac{q^2}{c^2},\lambda,\zh\right)&=&a_0\,\zh^4\,\Big(1+f_1\,\zh+f_2\,\zh^2+f_3\,
\zh^3+f_4\,\zh^4+{ O}(\zh^5)\Big)  \,\,\, .
\end{eqnarray}
The coefficients in \eqref{Kexp} read:
\begin{eqnarray}
A_2&=&0 \nn\\
B_2&=&\frac{1}{16}\frac{q^2}{c^2} \nn\\
C_2&=&-\frac{1}{16}\frac{q^2}{c^2}\lambda \\
D_2&=&\frac{7723}{2520000}\lambda^4-\frac{24127}{403200}\frac{q^2}{c^2}\lambda^2
-\frac{1079}{25200}\lambda^2  \non \\
&+&\frac{29}{4608}\frac{q^4}{c^4}+\frac{65}{1152}\frac{q^2}{c^2}+\frac{7}{144}\non\\
E2&=&\frac{1}{64}\Big(\frac{q^4}{c^4}+4\frac{q^2}{c^2}(1-2\lambda^2)\Big)
\,\,\, \nn
\end{eqnarray}
and
\begin{eqnarray}
f_1&=&\frac{4}{5}\lambda \nn\\
f_2&=&\frac{1}{12}\Big(\frac{q^2}{c^2}+8+4\,\lambda^2\Big) \nn\\
f_3&=&\frac{\lambda}{21}\Big(\frac{13}{10}\frac{q^2}{c^2}+12+2\,\lambda^2\Big) \\
f_4&=&\frac{1}{384}\left(\frac{q^4}{c^4}+\frac{q^2}{c^2}\left(20+\frac{92}{10}\lambda^2\right)+8\,\lambda^4+96\,\lambda^2+96\right)  \,\,\, .\nn
\end{eqnarray}
It is worth pointing out that the two expansions obtained by
this method do not correspond to the expansions, respectively,
of the functions $\wt{K}_1$ and  $\wt{K}_2$ in
\eqref{LaguerreTricomi1} and \eqref{LaguerreTricomi2} when
$\lambda=0$, but represent solutions in a different basis of
functions.

The AdS two-point correlation function:
\begin{equation}
\Pi^{(\lambda)}_{AdS}(q^2)=-\frac{R^3}{k}c^4\wt{K}\Big(\frac{q^2}{c^2},\lambda,\zh\Big)\frac{e^{-\zh^2-\lambda\,\zh}}{\zh^3}\partial_{\zh}\wt{K}\Big(\frac{q^2}{c^2},\lambda,\zh\Big)\Big|_{\zh
\rightarrow0}^{\zh \rightarrow +\infty}
\end{equation}
takes the form in the large space-like $q^2$ limit (regularizing
the $\zh \rightarrow +\infty$ contribution):
\begin{eqnarray}
\Pi^{(\lambda)}_{AdS}(q^2)&=&\frac{R^3}{k}\Big\{-q^4\Big[\frac{11}{288}+\frac{1}{8}\ln(c^2z^2)\Big]-q^2\Big[\frac{1}{2z^2}\Big(1-2\,c\,z\,\lambda\Big)\non \\
&&+\frac{47}{72}c^2\Big(1-\frac{8377}{16450}\lambda^2\Big)
+\frac{c^2}{2}\ln(c^2z^2)\Big(1-\frac{\lambda^2}{2}\Big)\Big] \nn \\
&&+c^4\Big[4
{B}^{\prime}-\frac{7}{9}\Big(1-\frac{1079}{1225}\lambda^2+\frac{7723}{122500}\lambda^4\Big)\Big]\Big\}
\,\, .
\end{eqnarray}
The logarithmic term in the perturbative contribution  remains
unchanged, and the parameter $k$ is still given by
eq.\eqref{kmatch}. This is an understandable result: the dilaton
does not modify the AdS dynamics in the $z\to0$ (UV) regime,
therefore duality with the perturbative QCD term is not affected
by its perturbations.

On the contrary, the other terms are modified, but  the effect
of perturbations on the condensate is quadratic in $\lambda$.
The dimension four gluon condensate is now given by
\begin{equation}\label{pert}
\langle \frac{\alpha_s}{ \pi }F^2\rangle =\frac{4
\alpha_s}{\pi^3} c^4 \left(4 {\eta}_0^{\prime}- \frac{7}{9}
\left(1-\frac{1079}{1225}\lambda^2+\frac{7723}{122500}\lambda^4\right)
\right) \,\,\ .
\end{equation}
The perturbations occur at ${O}(\lambda^2)$, showing that the
condensate is robust against perturbations of the dilaton field
of the form \eqref{dilatonpert}.

\section{Conclusions}
We have investigated duality between the QCD and the AdS
description of the two-point  correlation functions of the
scalar glueballs, finding that the QCD short-distance expansion
can be recovered in the AdS approach; however, the condensates
depend on a condition which, in addition to the normalization at
$z \to 0$, fixes the bulk-to-boundary propagator in the IR. We
have obtained  the spectra and decay constants of the scalar
glueballs.  Moreover, we have investigated how AdS can reproduce
a low energy QCD theorem for the two-point scalar glueball
correlation function. Assuming a particular form of the
coefficient function $\wt{B}$ which reproduces the low-energy
theorem, we have determined the dimension four gluon condensate,
which turns out to be positive, close to the commonly accepted
value and compatible with the AdS/QCD results obtained
investigating the two-point vector correlation function and the
Wilson loop.

\begin{acknowledgments}
We are indebted to H. Forkel for comments. One of us (FDF)
thanks Prof. A.J.Buras and the T31 group at Technische
Universit\"at, M\"unchen. This work was supported in part by the
EU contract No. MRTN-CT-2006-035482, "FLAVIAnet".
\end{acknowledgments}

\end{document}